\documentclass{elsart}
\usepackage{graphicx}
\usepackage{dcolumn}
\usepackage{bm}
\usepackage{amsmath}
\usepackage{amsfonts}
\usepackage{amssymb}

\begin{document}
\begin{frontmatter}
\title{Quantum Random Walk on the Line as a Markovian Process}
\author{A. Romanelli,}
\author{A.C. Sicardi Schifino\thanksref{dec}}
\author{, R. Siri,}
\author{G. Abal,}
\author{A. Auyuanet and R. Donangelo\thanksref{UFRJ}}
\address{Instituto de F\'{\i}sica, Facultad de Ingenier\'{\i}a\\
Universidad de la Rep\'ublica\\ C.C. 30, C.P. 11000, Montevideo, Uruguay}
\thanks[dec]{deceased}
\thanks[UFRJ]{Permanent address: Instituto de F\'{\i}sica, Universidade Federal do Rio de
Janeiro\\
C.P. 68528, 21941-972 Rio de Janeiro,Brazil}
\date{\today}
\begin{abstract}
We analyze in detail the discrete--time quantum walk on the line
by separating the quantum evolution equation into Markovian and
interference terms. As a result of this separation, it is possible
to show analytically that the quadratic increase in the variance
of the quantum walker's position with time is a direct consequence
of the coherence of the quantum evolution. If the evolution is
decoherent, as in the classical case, the variance is shown to
increase linearly with time, as expected.
Furthermore we show that this system has an evolution operator
analogous to that of a resonant quantum kicked rotor.
As this rotator may be described through a quantum computational
algorithm, one may employ this algorithm to describe the time
evolution of the quantum walker.
\end{abstract}
\end{frontmatter}

\section{Introduction}

\label{sec:introduction}

The study of computational devices based upon quantum mechanics, \textit{i.e.}
Quantum Computation, has drawn the attention of researchers in the last few
decades \cite{Feynman,Chuang}. The recent advances in technology that allow to
construct and preserve almost perfectly quantum states, have opened the
possibility of building useful quantum computing devices. However, relatively
few quantum algorithms that outperform classical ones have been found
\cite{Shor,Grover}. The classical random walk is an example of stochastic
motion that has found classical applications in many fields. The quantum
version of this problem has several features which are markedly different from
the classical walk \cite{Nayak,Childs}. As some classical algorithms are based
on random walks, it seems natural to ask whether quantum random walks might be
a useful tool for quantum computation \cite{Kempe}.

The discrete--time quantum walk on the line was introduced as a generalization
of the classical random walk to the quantum world. Here, we will focus on the
discrete--time quantum random walk on the line from a new perspective which
emphasizes the role of coherence as the physical reason behind the striking
differences found in the quantum version of the walk. We first briefly
introduce the basic notions and notation relative to the discrete--time
quantum walk on the line. Consider a particle that can move freely over a
series of interconnected sites. The discrete quantum walk on the line is
implemented by introducing an additional degree of freedom, the chirality,
which can take two values: ``left'' or ``right'', $|L\rangle$ or $|R\rangle$,
respectively. This is the quantum analog of the classical decision of the
random walker. At every time step, a rotation (or, more generally, a unitary
transformation) of the chirality takes place and the particle moves according
to its final chirality state. The global Hilbert space of the system is the
tensor product $H_{s}\otimes H_{c}$ where the Hilbert space associated to the
motion on the line is $H_{s}$ and the chirality Hilbert space is $H_{c}$.

If one is only interested in the properties of the probability distribution,
it has been claimed \cite{Nayak,Tregenna,Bach} that it suffices to consider
unitary transformations which can be expressed in terms of a single real
angular parameter $\theta$. Let us call the operators that translate the
walker one site to the left (right) on the line in $H_{s}$ as $T_{-}$ ($T_{+}%
$), and $|L\rangle\langle L|$ and $|R\rangle\langle R|$ the chirality
projector operators in $H_{c}$. We consider transformations of the form,
\begin{equation}
U(\theta)=\left\{  T_{-}\otimes|L\rangle\langle L|+T_{+}\otimes|R\rangle
\langle R|\right\}  \circ\left\{  I\otimes K(\theta)\right\}  , \label{Ugen}%
\end{equation}
where $K(\theta)=\sigma_{z} e^{i\theta\sigma_{y}}$, $I$ is the identity
operator in $H_{s}$, and $\sigma_{y}$ and $\sigma_{z}$ are Pauli matrices
acting in $H_{c}$. The unitary operator $U(\theta)$ evolves the state by one
time step,
\begin{equation}
|\Psi(t+1)\rangle=U(\theta)|\Psi(t)\rangle. \label{evol1}%
\end{equation}

One of the most remarkable characteristics of the quantum walk on the line is
that it spreads over the line faster than its classical counterpart. In this
work, we apply a general approach which leads to an physical insight of why
this is so. To do this, we rewrite the evolution equation as the sum of two
separate terms, one responsible for the classical--like diffusion and the
other for the quantum coherence \cite{Alejo}. As we shall see, the terms
responsible for the diffusion obey a master equation, as is typical of
Markovian processes, while the other includes the interference terms needed to
preserve the unitary character of the quantum evolution. In
Section~\ref{sec:derivation}, we review the decomposition of the evolution in
these two terms. In Section~\ref{sec:moments}, we obtain analytical
expressions for the first and second moments of the probability distribution
for these terms, in the case of the quantum random walker. Finally, in
Section~\ref{sec:conclusions}, we discuss our conclusions.

\section{Derivation of the Master Equation from the unitary evolution}

\label{sec:derivation}

In a recent work \cite{Alejo}, we have shown in detail how a unitary quantum
mechanical evolution can be separated into Markovian and interference terms.
This approach provides a new intuitive framework which proves useful for
analyzing the behavior of quantum systems in which decoherence plays a central
role. It is particularly suited to describe the evolution of quantum systems
which have classically diffusive counterparts.

The unitary evolution associated to $U(\theta)$ can be decomposed into a
Markovian term and an interference term. We begin by expressing the wave
vector, as the spinor
\begin{equation}
|\Psi(t)\rangle=\sum\limits_{m=-\infty}^{\infty}{\binom{a_{m}(t)}{b_{m}(t)}%
}|m\rangle, \label{spinor}%
\end{equation}
where we have associated the upper (lower) component to the left (right)
chirality and the states $|m\rangle$ are eigenstates of the position operator
corresponding to the site $m$ on the line. The unitary evolution for
$|\Psi(t)\rangle$, corresponding to eq.(\ref{evol1}), can then be written as
the map
\begin{align}
a_{i}(t+1)  &  =a_{i+1}(t)\,\cos\theta+b_{i+1}(t)\,\sin\theta\nonumber\\
b_{i}(t+1)  &  =a_{i-1}(t)\,\sin\theta-b_{i-1}(t)\,\cos\theta\,. \label{mapa}%
\end{align}
Note that for the particular case $\theta=\pi/4$, the usual Hadamard walk on
the line is obtained. The cases $\theta=0$ and $\theta=\pi/2$ are trivial
motions not considered here. We define the left-distribution of position as
$P_{mL}(t)~\equiv~\left|  a_{m}(t)\right|  ^{2}$ and the right-distribution as
$P_{mR}(t)~\equiv~\left|  b_{m}(t)\right|  ^{2}$. Then, the probability
distribution for the position is $P_{m}(t)\equiv P_{mL}(t)+P_{mR}(t)$ and
these distributions satisfy the map
\begin{align}
P_{i,L}(t+1)  &  =P_{i+1,L}(t)\,\cos^{2}\theta+P_{i+1,R}(t)\,\sin^{2}%
\theta+\beta_{i+1}(t) \,\sin2\theta\nonumber\\
P_{i,R}(t+1)  &  =P_{i-1,L}(t)\,\sin^{2}\theta+P_{i-1,R}(t)\,\cos^{2}\theta-
\beta_{i-1}(t) \,\sin2\theta\,. \label{mapa1}%
\end{align}
where the interference term in eq.(\ref{mapa1}) has been renamed as $\beta
_{i}\equiv\Re\left[  a_{i}b_{i}^{\ast}\right]  $ with $\Re(z)$ indicating the
real part of $z$.

It will prove useful to write equations (\ref{mapa1}) in the form%
\begin{equation}
P_{is}(t+1)=\mathop{\sum_{j=-\infty}}_{l=L,R}^{\infty}T_{ij,sl}P_{jl}%
(t)+\beta_{is}(t)\,\sin2\theta\label{master}%
\end{equation}
where $s,l$ take the two values of chirality $L,R$, $\beta_{iL}=\beta_{i+1}$
and $\beta_{iR}=-\beta_{i-1}$. The transition probabilities $T_{ij,sl}$ are
defined as
\begin{align}
T_{ii+1,LL}  &  =T_{ii-1,RR}=\cos^{2}\theta\nonumber\\
T_{ii+1,LR}  &  =T_{ii-1,RL}=\sin^{2}\theta\nonumber\\
T_{ij,ls}  &  =0\qquad\text{otherwise.} \label{eq:tprob}%
\end{align}
Note that these transition probabilities satisfy the necessary requirements
$T_{ij,sl}~\geq~0$ and $\sum_{i,s}T_{ij,sl}=\sum_{j,l}T_{ij,sl}=1$. Now it is
clear that if the interference term $\beta_{is}(t)$ in eq.(\ref{master}) can
be neglected, the time evolution of the occupation probability is described by
a Markovian process in which the transition probability $\left(  i,s\right)
\rightarrow\left(  j,l\right)  $ in a time $\Delta t=1$, is given by
$T_{ij,sl}$. As is characteristic of Markovian processes, the new position and
chirality depend only on the previous values for position and chirality.
Since the chirality is an auxiliary dimension introduced to implement the
quantum walk, we focus on the evolution of the position distribution for the
particle, $P_{i}=P_{iL}+P_{iR}$ , which can be obtained from eqs.(\ref{mapa1})
as,
\begin{align}
P_{i}(t+1)  &  =\left[  P_{i+1}(t)+P_{i-1}(t)\right]  \cos^{2}\theta
-P_{i}(t-1)\cos2\theta\nonumber\\
&  \qquad\qquad\qquad\qquad\qquad+\left[  \beta_{i+1}(t)-\beta_{i-1}%
(t)\right]  \sin2\theta. \label{probabilidad}%
\end{align}
We have used the fact that $P_{m}(t-1)=\left|  a_{m-1}(t)\right|  ^{2}+\left|
b_{m+1}(t)\right|  ^{2}$, a relation that is a consequence of the
map~(\ref{mapa}). Note that if the interference terms are neglected in
eq.(\ref{probabilidad}), the resulting evolution is Markovian in two time
steps. In the general case, eq.(\ref{probabilidad}) does not possess 
a continuum equivalent, even in the absence of interference terms.
However, there are two special cases where it is possible to obtain
continuum limits\cite{colin}. One is if we take the increments in space 
($\Delta x$) and the increments in time ($\Delta t$) in such a way that 
the velocity $v=\frac{\Delta x}{\Delta t}$ and the ratio
$\frac{\sin^{2}\theta}{\Delta t}$ are kept constant when $\Delta t\rightarrow 0$; 
in this case one obtains the Telegraphist's equation~\cite{goldstein}.
\begin{equation}
\frac{\partial P}{\partial t}=D\left[
\frac{\partial^{2}P}{\partial x^{2}}-\frac{\partial^{2}P}{\partial
t^{2}}\right]
\label{probabilidad1}
\end{equation}

In the second case, $v^{2}\Delta t$ is kept constant as $\Delta t\rightarrow 0$
and $\theta$ can take any value. Then the more familiar classical diffusion
equation is obtained.

\begin{equation}
\frac{\partial P}{\partial t}=D\frac{\partial^{2}%
P}{\partial x^{2}}
\label{probabilidad2}%
\end{equation}
where, in the previous two equations, $D$ is the diffusion coefficient,
\begin{equation}
D(\theta)=\frac{\cot^{2}\theta}{2}. \label{eq:D_coef}%
\end{equation}
Note that this expression reduces to the classical value, $D=1/2$, for the
Hadamard walk. The inclusion of the interference terms in the continuum
equations (\ref{probabilidad1},\ref{probabilidad2}) is done by adding
$2\cot\theta\frac{\partial\beta}{\partial x}$ to their right hand sides.

\section{Moments for the position distribution}

\label{sec:moments}

The evolution of the variance, $\sigma^{2}=M_{2}-M_{1}^{2}$, of the
distribution $P_{i}(t)$ is a distinctive feature of the quantum walk. It is
known \cite{travaglione} that it increases quadratically in time in the
quantum case, but only linearly in the classical case. We obtain the evolution
of the variance analytically from the evolution of the first and second
moments, defined as $M_{1}=\sum\limits_{i=-\infty}^{\infty}i\,P_{i}$ and
$M_{2}=\sum\limits_{i=-\infty}^{\infty}i^{2}P_{i}$ respectively. The evolution
equation for these moments is, from eq.(\ref{probabilidad}), written as%
\begin{align}
M_{1}(t+1)  &  = 2\cos^{2}\theta\, M_{1}(t)\,- \cos2\theta\, M_{1}(t-1)
-2\sin2\theta\,\sum_{i} \beta_{i}(t)\nonumber\\
M_{2}(t+1)  &  = 2\cos^{2}\theta\left[  1+M_{2}(t)\right]  \, -\cos2\theta\,
M_{2}(t-1)-4\sin2\theta\sum_{i} i\beta_{i}(t). \label{energia}%
\end{align}
In differential terms, these equations become,
\begin{align}
\frac{d^{2}M_{1}}{dt^{2}}+2\tan^{2}\theta\,\frac{dM_{1}}{dt}+4\tan\theta
\,\sum_{i}\beta_{i}(t)  &  =0\nonumber\\
\frac{d^{2}M_{2}}{dt^{2}}+2\tan^{2}\theta\,\frac{dM_{2}}{dt}+8\tan\theta\,
\sum_{i} i\beta_{i}(t)  &  =2 \label{energia1}%
\end{align}

\subsection{Decoherent evolution}

\label{ssec:decoherent}

If the sums in eqs.(\ref{energia1}) corresponding to the interference terms
can be neglected, the evolution of the moments of the distribution is given
by
\begin{align}
\frac{d^{2}M_{1}}{dt^{2}}+2\tan^{2}\theta\,\frac{dM_{1}}{dt}  &  =0\nonumber\\
\frac{d^{2}M_{2}}{dt^{2}}+2\tan^{2}\theta\,\frac{dM_{2}}{dt}  &  =2.
\label{energia2}%
\end{align}
The general solution for these equations is of the form
\begin{align}
M_{1}(t)  &  =C_{11}+C_{12}e{}^{-2t\tan^{2}\theta}\nonumber\\
M_{2}(t)  &  =C_{22}+\frac{1}{\tan^{2}\theta} \,t+C_{21}e^{-2t\tan^{2}\theta}
\label{eq:solm2}%
\end{align}
where $C_{11}$, $C_{12}$, $C_{21}$ and $C_{22}$ are constants. For times
larger than $\tau=\cot^{2}\theta$, the transient exponential is negligible and
the variance increases linearly in time with a slope given by
eq.(\ref{eq:D_coef}),
\begin{equation}
\sigma^{2}\approx D(\theta)\,t. \label{linear}%
\end{equation}
This is consistent with the expected result for a classical random walk on a
line. We emphasize that we have obtained this result neglecting the
interference terms in the exact quantum evolution, but this result is valid
for any $\theta$.

\subsection{Unitary evolution}

\label{ssec:hadamard}

\begin{center}
\begin{figure}[ptb]
\includegraphics[scale=0.8,angle=-90]{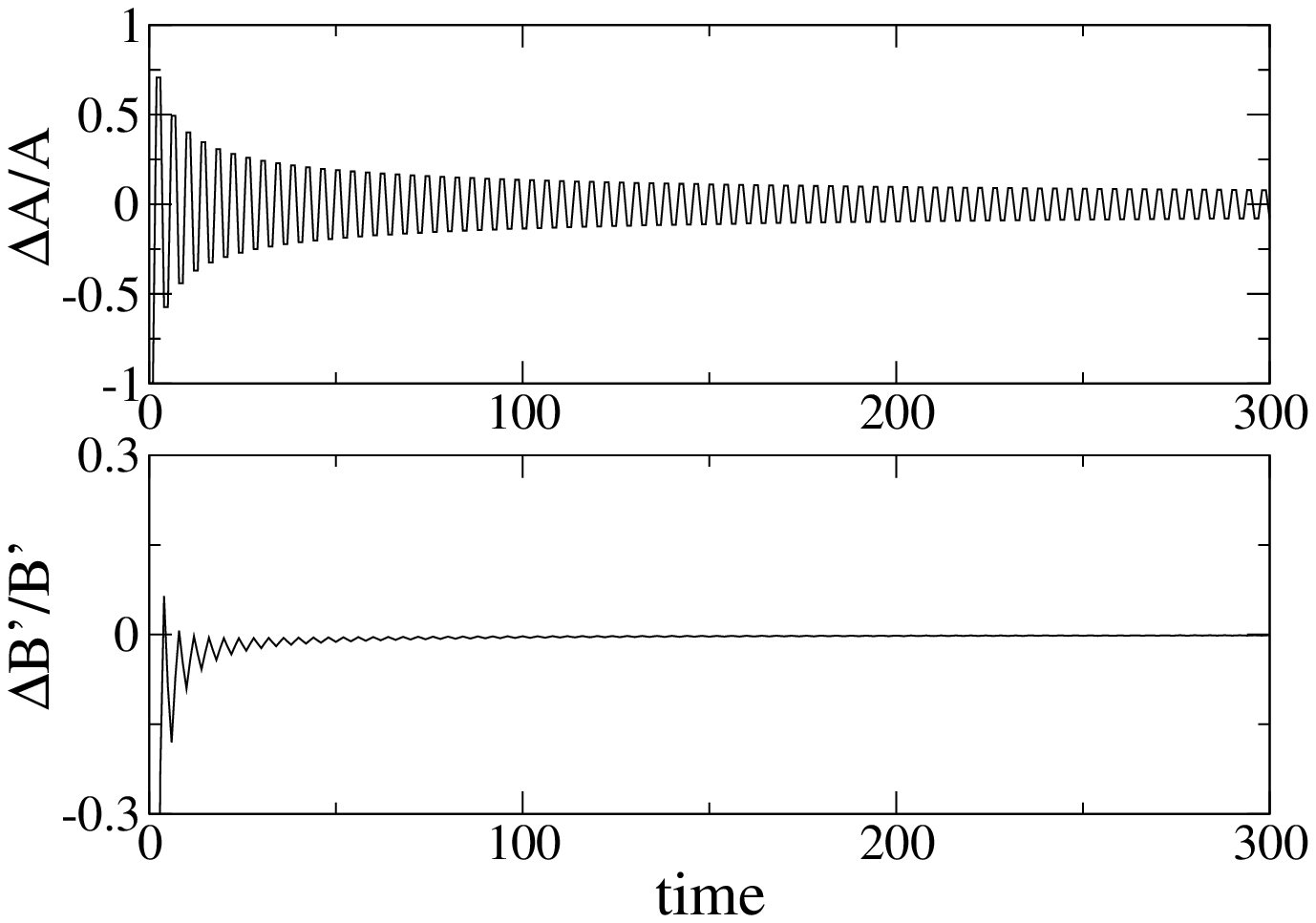}\caption{{\footnotesize The
fractional difference between the asymptotic value and the finite time value
of the expressions in eqs.(\ref{lim3}) and (\ref{lim4}). The quantities
$\Delta A$ and $\Delta B^{\prime}$ are defined in the text.}}%
\label{fig:f1}%
\end{figure}
\end{center}

If the interference terms in eqs.(\ref{energia1}) are not neglected, the
process is not Markovian but unitary. The solution for arbitrary $\theta$ is
cumbersome and not particularly illuminating. Therefore, in this subsection,
we particularize our solution for the case of the Hadamard walk setting
$\theta=\pi/4$ in the map (\ref{mapa}). Using Fourier analysis, the solutions
for the amplitudes $a_{i}(t)$ and $b_{i}(t)$ can be obtained \cite{Nayak}. For
the particular initial conditions $a_{j}(0)=\delta_{j0}$ and $b_{j}(0)=0$
these solutions are
\begin{align}
a_{j}(t)  &  =\frac{1+\left(  -1\right)  ^{j+t}}{2}\int_{-\pi}^{\pi}\frac
{dk}{2\pi}\left[  1+\frac{\cos k}{\sqrt{1+\cos^{2}k} }\right]
e^{-i\left(  w_{k}t+kj\right)  },\nonumber\\
b_{j}(t)  &  =\frac{1+\left(  -1\right)  ^{j+t}}{2}\int_{-\pi}^{\pi}\frac
{dk}{2\pi}\left[  \frac{e^{ik}}{\sqrt{1+\cos^{2}k}}\right]  e^{-i\left(
w_{k}t+kj\right)  }. \label{bes}%
\end{align}
where $\sin w_{k}=\frac{\sin k}{\sqrt{2}}$. From these amplitudes, an
expression for the variance can be obtained in the long time limit. The sums
in eqs.(\ref{energia1}) are evaluated, using the Fourier expansion of the
delta function, with the result
\begin{align}
\sum\limits_{j=-\infty}^{\infty}\beta_{j}(t)  &  =A\label{lim3}\\
\sum\limits_{j=-\infty}^{\infty}j\beta_{j}(t)  &  =-At+B. \label{lim4}%
\end{align}
The numerical constants are given by $A=\left(  2-\sqrt{2}\right)  /4$ and
$B=1-5\sqrt{2}/8$. Even though we have obtained these results in the long time
limit, for finite times they correctly describe the average trend. We have
checked this fact numerically, by computing the relative differences $\Delta
A/A\equiv\left[  \sum_{j}\beta_{j}-A\right]  /A$ and $\Delta B^{\prime
}/B^{\prime}\equiv\left[  \sum_{j} j\beta_{j} +At-B\right]  /(-At+B)$ as a
function of time. The result is shown in Fig.~\ref{fig:f1}.

Using eq.(\ref{energia}), the evolution for the first and the second moments
can be expressed as the maps
\begin{align}
M_{1}(t+1) &  =M_{1}(t)-2A\nonumber\\
M_{2}(t+1) &  =M_{2}(t)+4At+(1-4B).\label{energia3}%
\end{align}
The solutions for these maps are
\begin{align}
M_{1}(t) &  =-2At+C\nonumber\\
M_{2}(t) &  =2At^{2}+(1-4B-2A)t+C^{\prime},\label{sol2}%
\end{align}
with $C$ and $C^{\prime}$ arbitrary constants. The variance $\sigma^{2}%
=M_{2}-M_{1}^{2}$ is then
\begin{equation}
\sigma^{2}\approx2A(1-2A)\,t^{2}.\label{variance}%
\end{equation}
This result is consistent with previous those of previous works. In
ref.\cite{travaglione}, it is obtained through numerical simulations, in
\cite{Nayak} it is found through Fourier analysis, while in \cite{Konno}, the
same expressions are obtained through a summation over different paths.
However, it is important to emphasize that, from eqs.(\ref{lim3}) and
(\ref{variance}), a direct relation between a coherent evolution and the long
time quadratic increase of the variance has been established. In a decoherent
evolution $A$ and $B$ are negligible and the increase of the variance becomes
linear in time, as seen in eq.(\ref{sol2}). This is a particular instance of
the Markovian process discussed in the previous subsection.

\subsection{Generalized quantum walk}

\label{ssec:general} In this subsection we present an alternative analytical
approach which can be applied to the generalized quantum walk, i.e. for
arbitrary values of $\theta$. We start by rearranging the original map,
eqs.(\ref{mapa}), to uncouple both chirality components. The resulting
independent evolution equations are
\begin{align}
a_{i}(t+2)-a_{i}(t)  &  =\cos\theta\,\left[  a_{i+1}(t+1)-a_{i-1}(t+1)\right]
\nonumber\\
b_{i}(t+2)-b_{i}(t)  &  =\cos\theta\,\left[  b_{i+1}(t+1)-b_{i-1}(t+1)\right]
. \label{discreta2}%
\end{align}
For long times the left hand sides of the above equations can be approximated
by time derivatives. The result can be expressed as
\begin{equation}
2\frac{\partial\xi_{i}}{\partial t}=\cos\theta\left[  \xi_{i+1}(t)-\xi
_{i-1}(t)\right]  , \label{bessel1}%
\end{equation}
where $\xi_{i}$ stands for either chirality component, $a_{i}$ or $b_{i}$. We
define the effective time $t^{\prime}=-t\,cos\theta$, then eq.(\ref{bessel1})
becomes
\begin{equation}
2\frac{\partial\xi_{i}}{\partial t^{\prime}}=\xi_{i-1}(t^{\prime})-\xi
_{i+1}(t^{\prime}), \label{bessel2}%
\end{equation}
which is the recursion relation satisfied by the Bessel functions. Thus, we
can write the general solution as
\begin{equation}
\xi_{i}(t)=\sum_{-\infty}^{+\infty}(-1)^{i-l}\tilde{\xi}_{l}(0)J_{i-l}%
(t\,\cos\theta), \label{solucion1}%
\end{equation}
where $\tilde{\xi}_{l}(0)$ are the initial conditions to be used in the
differential equation (\ref{bessel2}). These initial conditions are not
necessarily the same as those to be used in the discrete map~(\ref{discreta2}%
), because the approximation of a difference by a continuous derivative does
not hold for small times. The solution (\ref{solucion1}) with appropriate
initial conditions provides a good long--time description of the dynamics of
the discrete map (\ref{discreta2}). In this context, long--time implies many
applications of the discrete map. Note that eq.(\ref{solucion1}) provides the
additional information \cite{Childs} that the long--time propagation speed of
the probability distribution is given by $\cos\theta$.

It is worth mentioning that the long--time solutions for the spinor
amplitudes, eq.(\ref{solucion1}), have the same form as the time-evolution of
the amplitudes of the kicked rotor in a principal resonance
\cite{izrailev,reso}. In fact, if $K$ is the dimensionless strength parameter
of the kicked rotor as defined in \cite{PRE_nuestro}, the time evolution for
these amplitudes is given by eq.(\ref{solucion1}) after the substitution
\begin{equation}
\cos\theta\rightarrow\frac{K}{4\pi p}\qquad\qquad\mbox{\textit{p}\;\;\;integer}.
\label{eq:K}%
\end{equation}

\begin{center}
\begin{figure}[ptb]
\includegraphics[angle=-90]{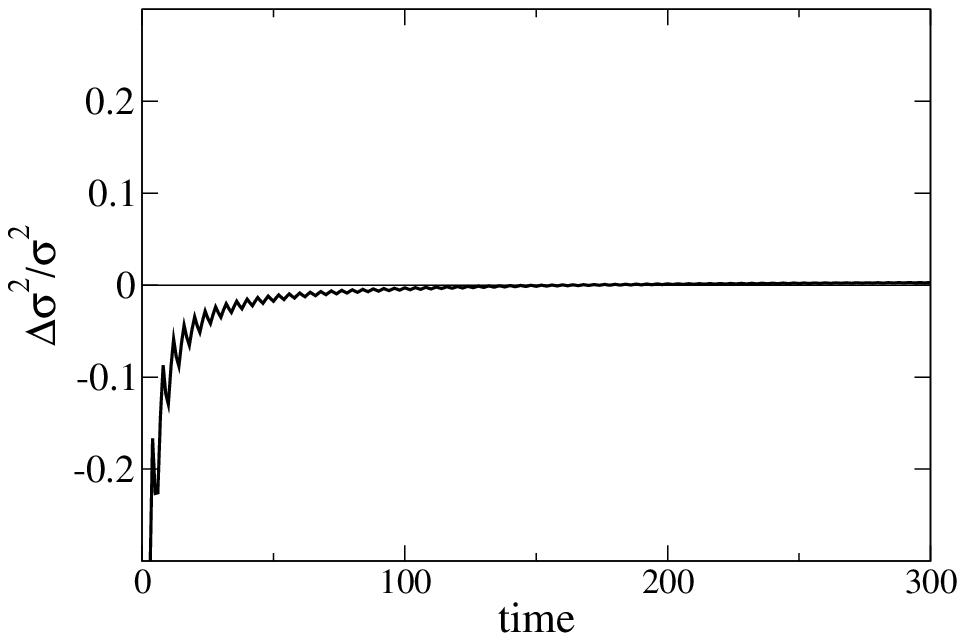}\caption{{\footnotesize Fractional
difference between the variance as obtained from the approximate expressions
(\ref{moment1}) and (\ref{moment2}), and the exact value $\sigma^{2}$ from the
original Hadamard map, eq.(\ref{mapa}). The quantity $\Delta\sigma^{2}$ is
defined in the text. The initial conditions used in eqs.(\ref{moment1}) and
(\ref{moment2}) are ${\binom{\tilde a_{0}}{\tilde b_{0}}}=0.70206{\binom{1}%
{1}}$, ${\binom{\tilde a_{\pm2}}{\tilde b_{\pm2}}}=-0.05963{\binom{1}{1}}$ and
zero otherwise. For the calculation of $\sigma^{2}$, the particle is initially
at the origin with chirality ${\binom{a_{0}}{b_{0}}}={\binom{1}{0}}$. }}%
\label{fig:f2}%
\end{figure}
\end{center}

We can obtain analytically the increase in the variance implied by
eq.(\ref{solucion1}). The position probability distribution can be expressed
as
\begin{equation}
P_{i}(t)=\sum_{l,l^{\prime}}(-1)^{-(l+l^{\prime})} \left[  \tilde
a_{l}(0)\tilde a_{l^{\prime}}^{\ast}(0)+\tilde b_{l}(0) \tilde b_{l^{\prime}%
}^{\ast}(0)\right]  \,J_{i-l}(t\cos\theta)\,J_{i-l}^{\prime}(t\cos\theta).
\label{distribucion}%
\end{equation}
The first and second moments of this distribution are
\begin{equation}
M_{1}(t)=-t\,\cos\theta\, \sum_{l} \Re\left[  \tilde a_{l}(0)\tilde
a_{l-1}^{\ast}(0) +\tilde b_{l}(0)\tilde b_{l-1}^{\ast}(0)\right]  + M_{1}(0)
\label{moment1}%
\end{equation}
and
\begin{align}
M_{2}(t)  &  =\frac{t^{2}\cos^{2}\theta}{2}\left\{  1+\sum_{l}\Re\left[
\tilde a_{l}(0)\tilde a_{l-2}^{\ast}(0)+ \tilde b_{l}(0)\tilde b_{l-2}^{\ast
}(0)\right]  \right\} \nonumber\\
&  \qquad-t\,\cos\theta\, \sum_{l} (2l-1)\Re\left[  \tilde a_{l}(0)\tilde
a_{l-1}^{\ast}(0) +\tilde b_{l}(0)\tilde b_{l-1}^{\ast}(0)\right]  + M_{2}(0).
\label{moment2}%
\end{align}
respectively, where $M_{1}(0)$ and $M_{2}(0)$ are the moments of the initial
distribution. Thus the variance $\sigma^{2}=M_{2}-M_{1}^{2}$ increases
quadratically in time for arbitrary initial conditions. This result holds for
arbitrary (non--trivial) values of the parameter $\theta$. In
Fig.~\ref{fig:f2} we show the time evolution of the relative difference
$\Delta\sigma^{2}/\sigma^{2}$ where $\Delta\sigma^{2}\equiv\sigma^{2}%
_{ap}-\sigma^{2}$ is the difference between the approximate variance
$\sigma^{2}_{ap}$ obtained from eqs.(\ref{moment1}) and (\ref{moment2}) and
the exact variance $\sigma^{2}$ from the original Hadamard map, eq.(\ref{mapa}).

The independence of the quadratic increase of $\sigma^{2}$ on the initial
conditions has been established numerically for the particular case of the
Hadamard walk \cite{travaglione}. Here we have also demonstrated the
independence of this quadratic increase on the parameter of the unitary
transformation. This is a new result.

\section{Conclusions}

\label{sec:conclusions}

The quadratic increase in time of the variance for the discrete time
generalized quantum walk has been obtained analytically. This increase remains
quadratic for arbitrary initial conditions and for all non--trivial values of
the parameter $\theta$ controlling the unitary transformation in the chirality
subspace. This quadratic increase results from interference effects and thus
it is strongly dependent on the coherence of the quantum evolution. If, for
any reason, the quantum evolution is decoherent then the increase in the
variance becomes linear with time.

The general approach of separating the quantum evolution equation into a
master equation supplemented by a term which takes into account quantum
coherence effects provides a new perspective which is helpful in clarifying
why the quantum evolution spreads faster than in the classical one. In the
quantum case, there is a superposition of a left-propagating wave and a
right-propagating wave, with both wavefronts traveling with constant speed
$\cos\theta$. Thus, their separation increases linearly in time and the
variance does so quadratically. In the Markovian approximation, as developed
in this work, the particle moves a step either to the right or to the left,
making its choice in a random way. Due to the randomness of this motion, the
variance increases only linearly with time. This process can be visualized as
a frequent position measurement process, which amounts to reinitialize the
system, at each measurement, in a different state. The wave function collapse
causes memory loss of the previous distribution. Thus, we have a Markovian
process which describes a series of random steps.

We have shown how the Markovian approximation method can be applied to a
generalized quantum walk, for arbitrary values of the parameter $\theta$ and
arbitrary initial conditions. When the evolution becomes decoherent, the
interference terms may be dropped and the evolution is described by a master
equation implying a linear increase of the variance with time. The diffusion
coefficient is not, in general, the same as that for the classical random
walk. Only in the particular case of the Hadamard walk the diffusion
coefficient is $1/2$ as in the classical case. This formalism shows in a
transparent form that the primary effect of decoherence is to make the
interference terms negligible in the evolution equation and then the Markovian
behavior is immediately implied. This is true for any evolution operator of
the form given in eq.(\ref{Ugen}).

We have established the analogy between the generalized quantum walk and the
resonant kicked rotor. This is done by obtaining an expression for the time
evolution of the chirality amplitudes of the quantum walk and showing that
they have the same form as the angular momentum components of the wavefunction
of the kicked rotor in a resonant regime. We related the kicked rotor strength
parameter to the parameter $\theta$ defining the unitary transformation of the
chirality. This analogy opens several interesting possibilities. A quantum
algorithm implements the evolution of the quantum kicked rotor in a quantum
computer is known \cite{Georgeot}. Since the quantum random walk on the line
has the same long--time dynamics as the resonant kicked rotor, the same
quantum algorithm may be used to describe the evolution of a generalized
discrete time quantum walk on the line. The quantum kicked rotor has been
experimentally realized using ultra-cold atom traps and some experiments have
focused on the resonant case \cite{exp_qkr}. This opens interesting
possibilities for the experimental realization of the quantum walk using
quantum optics.

Finally, note that the Markovian approach presented here provides a systematic
way to find the classical analog associated to a given quantum random walk. By
reformulating the quantum problem as described here and neglecting the
interference terms, the Markovian equation describing the equivalent classical
problem can be readily obtained.

We acknowledge the comments made by V. Micenmacher and the support of
\textit{PEDECIBA} and \textit{CONICYT (Clemente Estable proy. 6026) }. R.D.
acknowledges partial financial support from the Brazilian Research Council
(CNPq). A.R, G.A. and R.D. acknowledge financial support from the
\textit{Brazilian Millennium Institute for Quantum Information}--CNPq.

\end{document}